\documentclass[11pt,twoside,a4paper,cr,final]{cms-tdr}
\def\svnVersion{141193M}

\begin{document}

\hyphenation{had-ron-i-za-tion}
\hyphenation{cal-or-i-me-ter}
\hyphenation{de-vices}

\RCS$Revision: 138024 $
\RCS$HeadURL: svn+ssh://svn.cern.ch/reps/tdr2/papers/XXX-08-000/trunk/XXX-08-000.tex $
\RCS$Id: XXX-08-000.tex 138024 2012-07-19 04:04:00Z alverson $
\newlength\cmsFigWidth
\ifthenelse{\boolean{cms@external}}{\setlength\cmsFigWidth{0.85\columnwidth}}{\setlength\cmsFigWidth{0.4\textwidth}}
\ifthenelse{\boolean{cms@external}}{\providecommand{\cmsLeft}{top}}{\providecommand{\cmsLeft}{left}}
\ifthenelse{\boolean{cms@external}}{\providecommand{\cmsRight}{bottom}}{\providecommand{\cmsRight}{right}}
\cmsNoteHeader{-2012/163} 
\title{LHC Results on Charmonium in Heavy Ions}

\author*[ku]{Byungsik Hong on behalf of the ALICE, ATLAS, and CMS collaborations\\
Department of Physics, Korea University, Seoul 136-701, REPUBLIC OF KOREA}

\date{26 June 2012 (v2, 25 July 2012)}

\abstract{
In heavy-ion collisions at high energies, the quantum chromodynamics (QCD) predicts the production of the deconfined quark-gluon plasma (QGP) state. Quarkonia ($c\bar{c}$ or $b\bar{b}$ bound states) are a useful means to probe QGP and to investigate the behavior of QCD under the high parton-density environment. Up to now, the large hadron collider (LHC) at CERN provided two runs for PbPb collisions at $\sqrt{s_{NN}}$ = 2.76 TeV in the years 2010 and 2011. The ALICE, ATLAS, and CMS experiments at LHC have analyzed the yields and spectra of the $J/\psi$ and $\Upsilon$ families. In this article, we review particularly the recent charmonium results in PbPb collisions at LHC from the 2010 run. 
\\
\center{Presented at {\em Charm2012: The 5th International Workshop on Charm Physics}, University of Hawai'i, Honolulu, U.S.A., May 14-17, 2012}
}

\hypersetup{%
pdfauthor={CMS Collaboration},%
pdftitle={LHC Results on Charmonium in Heavy Ions},%
pdfsubject={CMS},%
pdfkeywords={CMS, physics}}

\maketitle 

\section{Introduction}\label{sec:intro}

QCD, the theory of strong interaction, predicts the deconfined quark-gluon plasma (QGP) state of partonic matter in high energy heavy-ion collisions. In particular, the heavy-ion collisions at colliders, such as LHC, create QGP in high temperature and small net baryon density, which resembles the primordial matter of the universe that existed about a few $\mu$s after the Big Bang. The temperature of such QGP should be at least 180 MeV, the transition temperature, with the energy density larger than 1 GeV/fm$^{3}$ \cite{qgp1}. 
  
\begin{figure}[hbt]
\centering
\includegraphics[width=8cm]{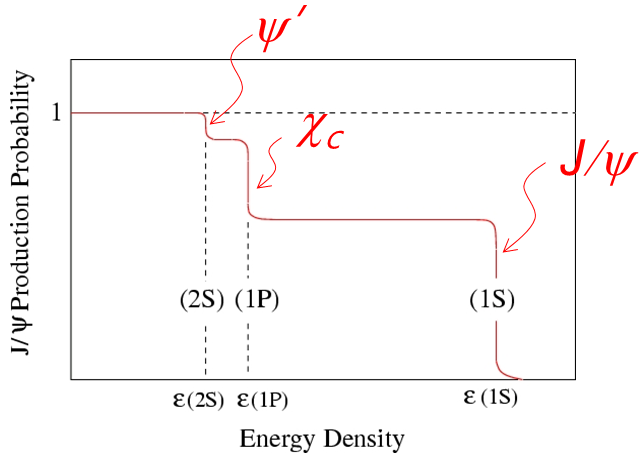}
\caption{Sequential suppression pattern of the $c\bar{c}$ states expected in high-energy heavy-ion collisions \cite{satz1}.}
\label{fig:sequence}
\end{figure}

Among several experimental signatures for QGP, the suppression of quarkonium states is one of the most prominent observables \cite{matsui1}. In the original picture, the number of quarkonia, created in primary nucleon-nucleon ($NN$) collisions, should be significantly reduced by the dissociation of $Q \bar{Q}$ ($Q$ = $c$ or $b$) bound states during the deconfined phase of quarks and gluons due to the color screening of the long-range QCD confining potential. Various quarkonium states are expected to be sequentially dissolved due to the color screening effect, depending on their binding energies and the temperature of the surrounding matter \cite{satz1,satz2}. As a result, quarkonia can play the role of a thermometer for the temperature reached by heavy-ion collisions and provide valuable information on the detailed properties of QCD in the high-density phase \cite{mocsy1}. Table.~\ref{tab:qqbar} summarizes the properties of $J/\psi$(1$S$), $\chi_{c}$(1$P$) and $\psi'$(2$S$) obtained by the non-relativistic potential model \cite{satz1,satz2}. Since the binding energies and radii are different for various $c\bar{c}$ states, a sequential suppression pattern similar to Fig.~\ref{fig:sequence} is expected \cite{satz1}. 

\begin{table}[h]
\begin{center}
\begin{tabular}{l|ccc}  
State & $J/\psi$(1$S$) & $\chi_{c}$(1$P$) & $\psi'$(2$S$) \\ \hline
Mass (GeV/$c^{2}$) & 3.10 & 3.53 & 3.68 \\
Binding energy (GeV) & 0.64 & 0.20 &0.05 \\ 
Radius (fm) & 0.25 & 0.36 & 0.45 \\ \hline
\end{tabular}
\caption{Masses, binding energies, and radii of various charmonium states from non-relativistic potential model \cite{satz1,satz2}.}
\label{tab:qqbar}
\end{center}
\end{table}

\begin{figure}[t!]
\centering
\includegraphics[width=7.2cm]{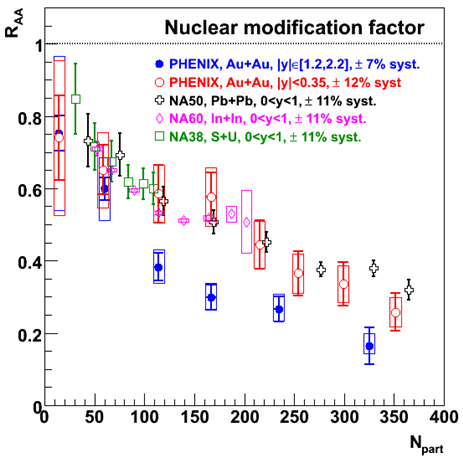}
\caption{Comparison of the nuclear modification factor of $J/\psi$ as a function of the event centrality between the SPS ($\sqrt{s_{NN}}$ = 17 GeV) and RHIC ($\sqrt{s_{NN}}$ = 200 GeV) data \cite{sps1,sps2,sps3,phenix2}. At RHIC, the data at midrapidity (open circles) and forward rapidity (closed circles) are displayed. The nuclear modification factor $R_{AA}$ is the particle yield in nuclear ($AA$) collisions relative to that in $pp$ collisions after proper scaling by the number of binary collisions. (The definition of $R_{AA}$ will be described in detail in Sec.~\ref{sec:cms}.) $N_{\mathrm {part}}$ is the number of participating nucleons determined by the Glauber model so that a larger $N_{\mathrm {part}}$ means more central collisions with a smaller impact parameter.}
\label{fig:sps_rhic}
\end{figure}
 
However, there are some complications in interpreting the quarkonium production data in not only heavy-ion collisions, but also elementary $pp$ collisions. First of all, at present, there is no satisfactory model that can successfully describe both the cross section and the polarization of quarkonia even in $pp$ collisions. Secondly, no detailed information on the cold nuclear matter effects, like shadowing and gluon saturation in initial states, are available with enough precision \cite{phenix1}. Thirdly, a similar $J/\psi$ suppression at RHIC ($\sqrt{s_{NN}}$ = 200 GeV) and SPS ($\sqrt{s_{NN}}$ = 17 GeV) and a larger suppression at forward rapidity than at midrapidity in AuAu collisions at RHIC still remain as puzzles (Fig.~\ref{fig:sps_rhic}) \cite{sps1,sps2,sps3,phenix2}. Finally, the model calculations are not converging to the measured $J/\psi$ productions in nuclear collisions. For example, the statistical coalescence (or recombination or regeneration) model of uncorrelated $c$ and $\bar{c}$ quarks predicts a slight increase, but the holographic QCD (AdS/CFT) calculation with $J/\psi$s embedded in a hydrodynamic model predicts a large decrease of the $J/\psi$ yield with respect to the $pp$ data for the transverse momentum ($p_{T}$) larger than $\sim$5 GeV/$c$ in CuCu collisions at RHIC \cite{star1,phenix3}. As a result, the quarkonium production data at much higher LHC energies must be critical to resolve those questions. 
 
\begin{figure}[t!]
\centering
\includegraphics[width=10cm]{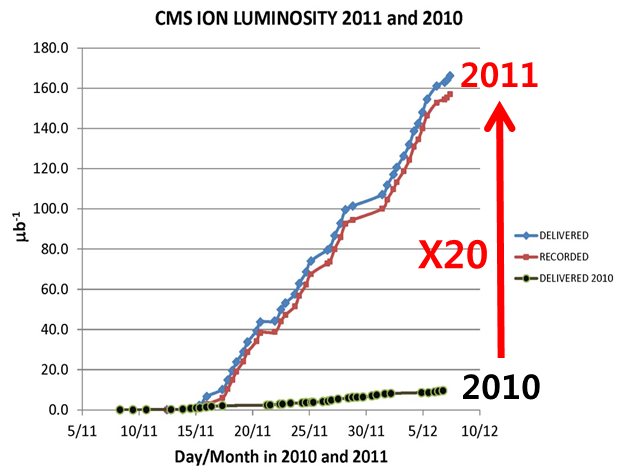}
\caption{Integrated luminosities delivered by LHC (top) and recorded by CMS (middle) in 2011 as a function of Day/Month. For comparison, the integrated luminosities for the first heavy-ion run in 2010 are also displayed in the bottom. About a factor of twenty enhancement was achieved in the integrated luminosities at LHC. }
\label{fig:int_lum}
\end{figure}

So far, LHC has provided PbPb collisions at $\sqrt{s_{NN}}$ = 2.76 TeV in two periods. During the first heavy-ion run in 2010, LHC delivered about 7 $\mu$b$^{-1}$ of integrated luminosity. In the following year 2011, thanks to the great improvement in the accelerator performance, LHC delivered about a factor of twenty more (about 150 $\mu$b$^{-1}$) PbPb events during a similar data taking period as 2010 (Fig.~\ref{fig:int_lum}). With much higher statistics of the new data sample accumulated in 2011, the detailed analyses for $\psi'$(2$S$) and higher $\Upsilon$ states become possible, but in this article, we present the experimental data only from the 2010 run. Note that LHC has also delivered $pp$ collisions at the same energy as PbPb in March 2011, and the experiments used them as references when they analyzed the PbPb data. For example, CMS has accumulated about 225 $\mu$b$^{-1}$ of the integrated luminosity of $pp$ ($\mathcal{L}_{pp}$), which is roughly the same number of binary collision scaled PbPb data recorded in 2010. 
 
This paper summarizes the charmonium production in PbPb collisions at $\sqrt{s_{NN}}$ = 2.76 TeV from the ALICE \cite{alice2}, ATLAS \cite{atlas1}, and CMS \cite{cms1} collaborations at LHC. The results on non-prompt $J/\psi$ from B-meson decay \cite{cms1} and $\Upsilon$ states \cite{cms-u} from CMS are also briefly presented. The detailed descriptions on the ALICE, ATLAS, and CMS detector systems can be found in \cite{alice-d,atlas-d,cms-d}. 

\section{$J/\psi$ at High-$p_{T}$ from ATLAS}\label{sec:atlas}

\begin{figure}[b!]
\centering
\includegraphics[width=9cm]{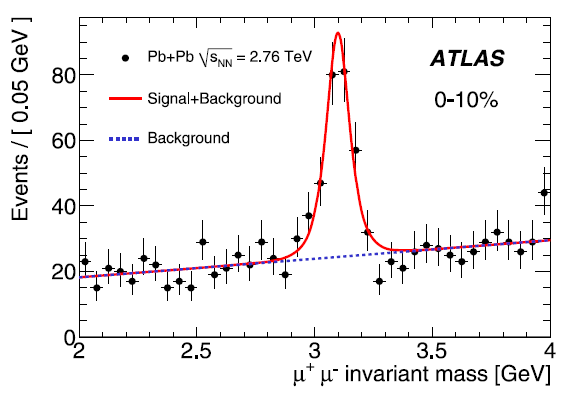}
\caption{Invariant mass distribution of $\mu^{+}\mu^{-}$ pairs in the $J/\psi$ region for the most central bin (0 - 10\%) in PbPb collisions at $\sqrt{s_{NN}}$ = 2.76 TeV from ATLAS \cite{atlas1}.}
\label{fig:atlas_mass}
\end{figure}

Soon after the first heavy-ion run at LHC, the ATLAS collaboration has deliberately published the centrality dependence of the $J/\psi$ yield \cite{atlas1}. Figure~\ref{fig:atlas_mass} displays the invariant mass distribution of $\mu^{+}\mu^{-}$ pairs in the $J/\psi$ region for the most central class (0 - 10\%) of PbPb collisions. Note that about 80\% of the $J/\psi$s reconstructed by ATLAS are in $p_{T} >$ 6.5 GeV/$c$. 

Since the reference $pp$ data were not available at the time of this analysis, ATLAS estimated the relative $J/\psi$ yield in each centrality bin `$cent$' by normalizing the yield in the most peripheral class (40 - 80\%), $N_{\mathrm{PbPb}}^{cent}(J/\psi)/N_{\mathrm{PbPb}}^{40-80\%}(J/\psi)$. Following this, the central-to-peripheral ratio, $R_{cp}$, was obtained by scaling the relative yield by the ratio of the mean number of binary collisions, $<N_{coll}^{cent}>/<N_{coll}^{40-80\%}>$: 
\begin{equation}
R_{cp} = \frac{N_{\mathrm{PbPb}}^{cent}(J/\psi)/<N_{coll}^{cent}>}
{N_{\mathrm{PbPb}}^{40-80\%}(J/\psi)/<N_{coll}^{40-80\%}>}. 
\label{eq:rcp}
\end{equation}

\begin{figure}[t!]
\centering
\includegraphics[width=8cm]{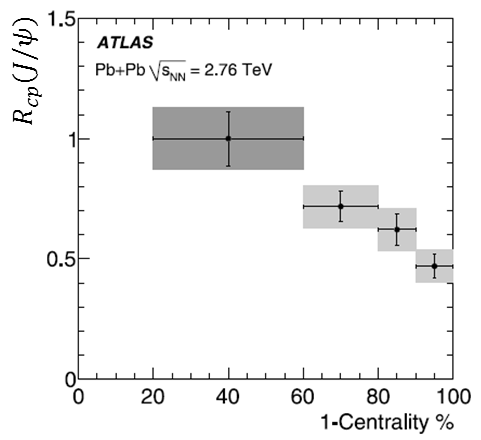}
\caption{$R_{cp}$ of $J/\psi$ as a function of centrality in PbPb collisions at $\sqrt{s_{NN}}$ = 2.76 TeV from ATLAS \cite{atlas1}. The vertical bars represent the statistical errors while the grey boxes represent the combined systematic errors.}
\label{fig:atlas_rcp}
\end{figure}

Figure~\ref{fig:atlas_rcp} summarizes $R_{cp}$ as a function of centrality. It clearly demonstrates that the yield of $J/\psi$ is suppressed for more central collisions with a smaller impact parameter. Note that the uncertainty displayed with the most peripheral point, which is equal to unity by construction, is global to all points. 

\section{$J/\psi$ at High-$p_{T}$ from CMS}\label{sec:cms}
 
The geometrical acceptance of the CMS detector is similar to that of the ATLAS detector for muons. The low $p_{T}$ limits of CMS for the accepted muon pairs are 6.5 GeV/$c$ for the midrapidity bin ($|y| <$ 1.2) and 3 GeV/$c$ for the most forward rapidity bin (1.6 $< |y| <$ 2.4) \cite{cms1}. CMS used the $pp$ data as references and, instead of $R_{cp}$, analyzed the nuclear modification factor $R_{AA}$: 
\begin{equation}
R_{AA} = {{{\mathcal L}_{pp}}\over{T_{AA}N_{MB}}}
{{N_{\mathrm {PbPb}}^{cent}(J/\psi)}\over{N_{pp}(J/\psi)}}
{{\epsilon_{pp}}\over{\epsilon_{\mathrm {PbPb}}}},
\label{eq:raa}
\end{equation}
where $T_{AA}$ is the nuclear overlap function estimated by the Glauber model, $N_{MB}$ is the count of equivalent minimum bias events in PbPb, and $\epsilon_{pp}/\epsilon_{\mathrm {PbPb}}$ is the multiplicity dependent ratio of the efficiencies in $pp$ and PbPb collisions for trigger and reconstruction. The ratio, $\epsilon_{pp}/\epsilon_{\mathrm {PbPb}}$, was determined by the Monte-Carlo simulation that embedded a PYTHIA signal event \cite{pythia} to a HYDJET background event \cite{hydjet}. In the simulation, $\epsilon_{pp}/\epsilon_{\mathrm {PbPb}}$ was estimated to be about 1.17 for the most central bin. By definition, deviation of Eq.~\ref{eq:raa} from unity indicates the medium effect on the $J/\psi$ production in heavy-ion collisions. 
 
\begin{figure}[t!]
\begin{center}
\includegraphics[width=14cm]{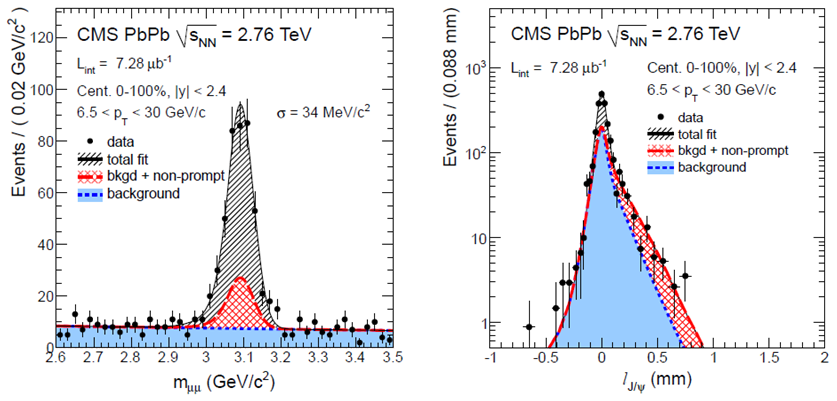}
\caption[]{
Invariant mass (left) and pseudo-proper decay length (right) distributions of $\mu^{+}\mu^{-}$ pairs for $|y| <$ 2.4 in minimum bias PbPb collisions at $\sqrt{s_{NN}}$ = 2.76 TeV from CMS \cite{cms1}. In the left panel, the solid line represents the sum of the `crystal ball' function for the signal and an exponential function for the background. In both panels, the dotted and dashed lines represent the backgound and the non-prompt $J/\psi$ component, respectively. The solid lines are the total fit functions including the prompt $J/\psi$ component.} 
\label{fig:mass_ljpsi}
\end{center}
\end{figure}

In addition, CMS classified the measured $J/\psi$s into the prompt component associated with the primaty vertex and the non-prompt component associated with a secondary vertex of $\mu^{+}\mu^{-}$ \cite{cms1}. The prompt and non-prompt components have different origins: the former is from either direct production or feed down of the higher states, such as $\psi^{\prime}$(2$S$) and $\chi_{c}$(1$P$), and the latter is from the B-meson decay. Because B mesons fly finite pathlength before they decay, the non-prompt $J/\psi$s can be separated from the prompt ones by the pseudo-proper decay length:
\begin{equation}
l_{J/\psi} = L_{xy} \frac{m_{J/\psi}}{p_{T}},
\label{eq:ljpsi}
\end{equation} 
where $L_{xy}$ is the distance between the primary and the secondary vertices in the transverse plane and $m_{J/\psi}$ is the rest mass of $J/\psi$. CMS has fitted the invariant mass and the pseudo-proper decay length distributions simultaneously to determine the yields of the prompt and non-prompt components. Figure~\ref{fig:mass_ljpsi} shows the invariant mass and the projected $l_{J/\psi}$ distributions of $\mu^{+}\mu^{-}$ pairs in PbPb collisions at $\sqrt{s}$ = 2.76 TeV \cite{cms1}. Note that the resolution of the $J/\psi$ peak is $\sim$34 MeV/$c^{2}$, which is compatible to that in $pp$ collisions.

\begin{figure}[t!]
\begin{center}
\includegraphics[width=8cm]{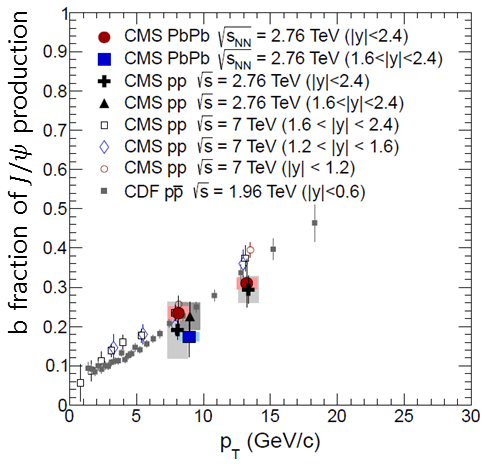}
\caption[]{Fraction of the B-decay contribution to the inclusive $J/\psi$ yield as a function of $p_{T}$ in $pp$ and PbPb collisions at $\sqrt{s_{NN}}$ = 2.76 TeV from CMS \cite{cms1}. For comparison, the $b$ fraction measured by CDF in $p\bar{p}$ collisions at $\sqrt{s}$ = 1.96 TeV \cite{cdf1} and by CMS in $pp$ collisions at $\sqrt{s}$ = 7 TeV \cite{cms2} are also displayed. Points are plotted at the measured average values of $p_{T}$. The statistical and systematic uncertainties are shown by bars and boxes, respectively.} 
\label{fig:b_frac}
\end{center}
\end{figure}

The fractions of B-meson decays in inclusive $J/\psi$s in $pp$ as well as PbPb collisions at $\sqrt{s_{NN}}$ = 2.76 TeV are presented in Fig.~\ref{fig:b_frac} as a function of $p_{T}$, integrated over centrality. For comparison, Fig.~\ref{fig:b_frac} also includes the data for $b$ fraction measured by CDF in $p\bar{p}$ at $\sqrt{s}$ = 1.96 TeV \cite{cdf1} and CMS in $pp$ at $\sqrt{s}$ = 7 TeV \cite{cms2}. Within uncertainties, a good agreement between the present data and earlier results can be found. 

\begin{figure}[t!]
\begin{center}
\includegraphics[width=8cm]{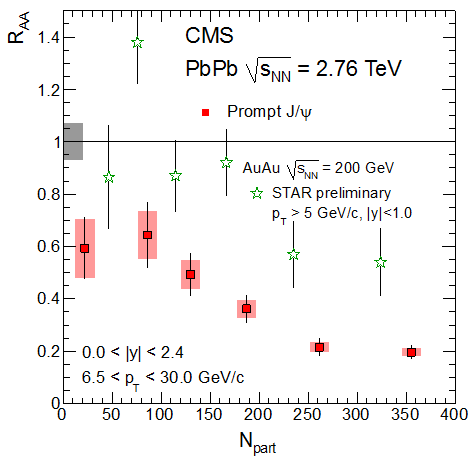}
\caption[]{
Nuclear modification factor $R_{AA}$ of prompt $J/\psi$ as a function of $N_{\mathrm{part}}$ in PbPb collisions at $\sqrt{s_{NN}}$ = 2.76 TeV from CMS \cite{cms1}. The measured $J/\psi$s are for $p_{T}$ from 6.5 to 30 GeV/$c$ in $|y| <$ 2.4. The grey box at $R_{AA}$ = 1 represents a global scale uncertainty of 6\% originating from the integrated luminosity of the $pp$ data sample. The statistical and systematic uncertainties are represented by bars and boxes, respectively. The CMS data are compared to the STAR data (stars) for $p_{T} >$ 5 GeV/$c$ and $|y| <$ 1.0 in AuAu collisions at $\sqrt{s_{NN}}$ = 200 GeV \cite{star2}.} 
\label{fig:raa_npart}
\end{center}
\end{figure}

The nuclear modification factor $R_{AA}$ of prompt $J/\psi$ is displayed in Fig.~\ref{fig:raa_npart} as a function of the participating nucleons $N_{\mathrm {part}}$ \cite{cms1}. $R_{AA}$ in Fig.~\ref{fig:raa_npart} is integrated over 6.5 $< p_{T} <$ 30.0 GeV/$c$ in $|y| <$ 2.4, and a large dependence on the collision centrality is observed. $R_{AA}$ of prompt $J/\psi$ is about 0.6 for 50 - 100\% centrality bin, and decreases to 0.2 $\pm$ 0.03(stat.) $\pm$ 0.01(syst.) $\pm$ 0.01(global), which indicates a factor of five suppression relative to $pp$, for the most central bin (0 - 10\%). 

Figure~\ref{fig:raa_npart} compares the current CMS results with the STAR preliminary data at RHIC ($p_{T} >$ 5 GeV/$c$ in $|y| <$ 1.0) measured in AuAu collisions at $\sqrt{s_{NN}}$ = 200 GeV \cite{star2}. This comparison reveals a larger suppression at higher beam energy at LHC for all centrality bins. Furthermore, CMS finds that the $J/\psi$ suppression apparently exists even in the most peripheral bin, which could not be observed in the $R_{cp}$ analysis done by ATLAS. 

\begin{figure}[t!]
\begin{center}
\includegraphics[width=14cm]{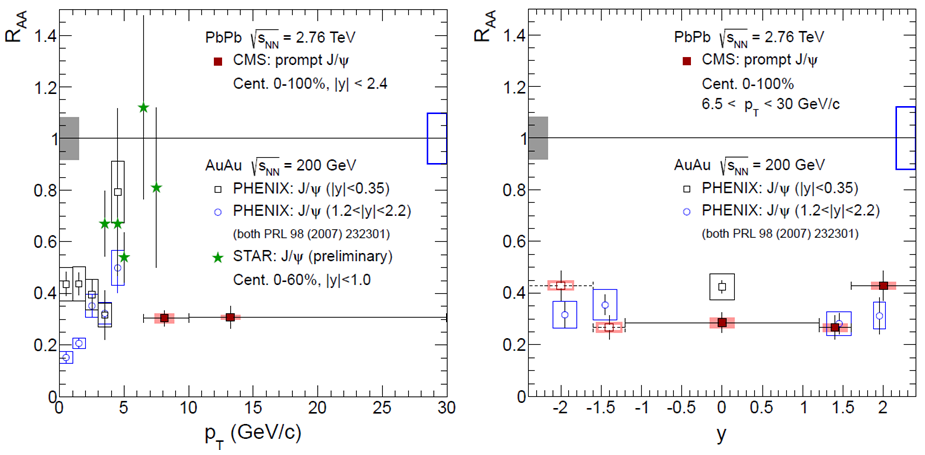}
\caption[]{
Nuclear modification factor $R_{AA}$ of prompt $J/\psi$, integrated over centrality, as functions of $p_{T}$ (left) and $y$ (right) in PbPb collisions at $\sqrt{s_{NN}}$ = 2.76 TeV from CMS \cite{cms1}. $R_{AA}$ is integrated over $|y| <$ 2.4 for the $p_{T}$ dependence in the left plot and over 6.5 $< p_{T} <$ 30.0 GeV/$c$ for the $y$ dependence in the right plot under the minimum-bias centrality condition. The grey boxes at $R_{AA}$ = 1 represent global scale uncertainties of 6\% originating from the integrated luminosity of the $pp$ data sample. The statistical and systematic uncertainties are shown by bars and boxes, respectively. The PHENIX and STAR data obtained at RHIC are also included for comparison \cite{star2,phenix-pty}.} 
\label{fig:raa_pty}
\end{center}
\end{figure}

Figure~\ref{fig:raa_pty} shows the $p_{T}$ and rapidity dependences of $R_{AA}$ for prompt $J/\psi$ in PbPb collisions at $\sqrt{s_{NN}}$ = 2.76 TeV from CMS, integrated over 0 - 100\% of centrality \cite{cms1}. Note that the data in Fig.~\ref{fig:raa_pty} is integrated over $|y| <$ 2.4 for the $p_{T}$ dependence in the left panel and over 6.5 $< p_{T} <$ 30.0 GeV/$c$ for the $y$ dependence in the right panel. Suppression by a factor of three is observed for two $p_{T}$ bins when integrated over centrality and rapidity without particular observable $p_{T}$ dependence. For the $p_{T}$ dependence of $R_{AA}$ the CMS data show a larger suppression at LHC energy than RHIC energy for $p_{T} >$ 5 GeV/$c$ \cite{phenix-pty}.

In the rapidity dependence of Fig.~\ref{fig:raa_pty}, the CMS data show that the yield of prompt $J/\psi$ is less suppressed at forward rapidity than midrapidity in $p_{T} >$ 6.5 GeV/$c$. This trend is opposite to what PHENIX has observed at the lower $p_{T}$ region ($<$ 4 GeV/$c$) as shown in Fig.~\ref{fig:sps_rhic} \cite{phenix-pty}. The difference between the CMS and the PHENIX data may come from (anti-)shadowing for different $x$ ranges covered and/or different contributions by the regenerated $J/\psi$s. However, further investigation is needed for the quantitative interpretation of the experimental data.

\begin{figure}[t!]
\begin{center}
\includegraphics[width=8cm]{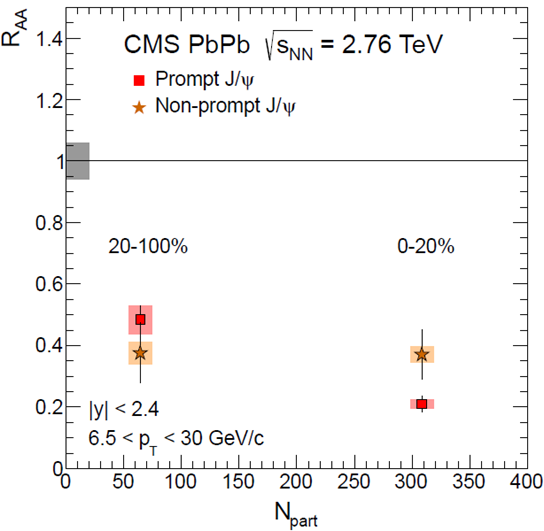}
\caption[]{
Nuclear modification factor $R_{AA}$ of non-prompt $J/\psi$ as a function of $N_{\mathrm{part}}$ in PbPb collisions at $\sqrt{s_{NN}}$ = 2.76 TeV from CMS \cite{cms1}. The $p_{T}$ range of $J/\psi$ is from 6.5 to 30.0 GeV/$c$ in $|y| <$ 2.4. The grey box at $R_{AA}$ = 1 represents a global scale uncertainty of 6\% originating from the integrated luminosity of the $pp$ data sample. The statistical and systematic uncertainties are shown by bars and boxes, respectively. For comparison, the $R_{AA}$ data of prompt $J/\psi$ obtained by CMS are also included after rebinned as non-prompt $J/\psi$ in $N_{\mathrm{part}}$.} 
\label{fig:nonprompt}
\end{center}
\end{figure}

Figure~\ref{fig:nonprompt} displays the nuclear modification factor $R_{AA}$ of non-prompt $J/\psi$ as a function of $N_{\mathrm {part}}$ in PbPb collisions at $\sqrt{s_{NN}}$ = 2.76 TeV \cite{cms1}. It shows that $R_{AA}$ of non-prompt $J/\psi$ is 0.37 $\pm$ 0.08(stat.) $\pm$ 0.02(syst.) $\pm$ 0.02(global) for the most central bin (0 - 20\%) in PbPb collisions, which implies the suppression by a factor of $\sim$2.7 relative to the $pp$ data. This is the first data hinting the $b$-quark energy loss in medium. No centrality dependence is noticed with the present wide bin size, especially, in the peripheral bin of 20 - 100\%. The data with finer centrality bins are expected from the ongoing analysis using higher statistics data sample obtained in 2011. 

\section{$J/\psi$ at Low-$p_{T}$ from ALICE}\label{sec:alice}
 
Because the muon spectrometer is positioned in a very forward region, ALICE can measure single muons down to $p_{T}$ = 0.5 GeV/$c$ and $J/\psi$s down to $p_{T}$ = 0 GeV/$c$ in 2.5 $< y <$ 4 \cite{alice-d}. Therefore, the geometrical acceptance of ALICE is quite complimentary to those of ATLAS and CMS. 
 
\begin{figure}[t!]
\centering
\includegraphics[width=15cm]{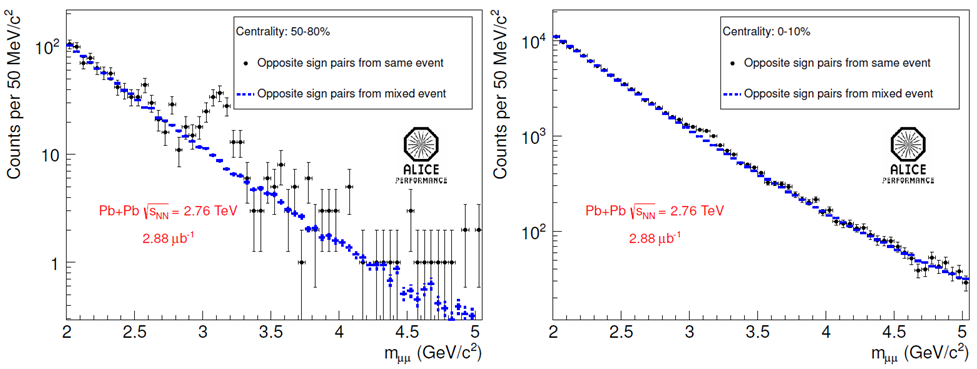}
\caption{Invariant mass distributions of $\mu^{+}\mu^{-}$ pairs from the same and mixed events in the $J/\psi$ region from ALICE \cite{alice1}. The left and right panels represent the semi-central (50 - 80\%) and the most central (0 - 10\%) classes, respectively, in PbPb collisions at $\sqrt{s_{NN}}$ = 2.76 TeV.}
\label{fig:alice_mass}
\end{figure}

Figure~\ref{fig:alice_mass} shows the invariant mass distributions of $\mu^{+}\mu^{-}$ pairs from the same events and the mixed events in the $J/\psi$ region \cite{pillot1,alice1}. Here, the invariant mass distributions from the same events include the $J/\psi$ signal as well as the backgrounds whereas those from the mixed events provide only the shape of the backgrounds. In particular, Fig.~\ref{fig:alice_mass} compares the invariant mass distributions for the semi-central class (50 - 80\%) with that for the most central class (0 - 10\%) in PbPb collisions at $\sqrt{s_{NN}}$ = 2.76 TeV from ALICE. As expected, the signal-to-background ratio is significantly reduced for more central collisions.

\begin{figure}[t!]
\begin{center}
\includegraphics[width=9cm]{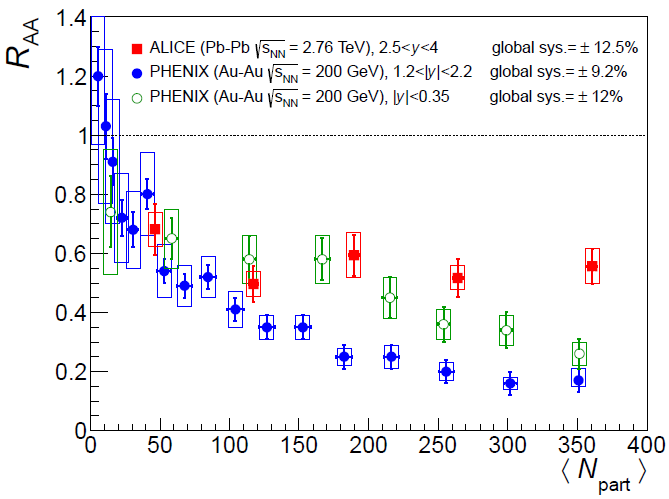}
\caption[]{
Nuclear modification factor $R_{AA}$ of inclusive $J/\psi$ as a function of $N_{\mathrm{part}}$ for $p_{T} \geq$ 0 GeV/$c$ and 2.5 $< y <$ 4 in PbPb collisions at $\sqrt{s_{NN}}$ = 2.76 TeV from ALICE \cite{alice2}. The statistical and systematic uncertainties are shown by bars and boxes, respectively. The ALICE data are compared to the PHENIX data (the open circles for midrapidity in $|y| <$ 0.35 and the filled circles for the forward rapidity in 1.2 $< |y| <$ 2.2) measured at low $p_{T}$ in AuAu collisions at $\sqrt{s_{NN}}$ = 200 GeV \cite{phenix2}.} 
\label{fig:alice_raa}
\end{center}
\end{figure}

Figure~\ref{fig:alice_raa} shows the nuclear modification factor $R_{AA}$ for inclusive $J/\psi$ in 2.5 $< y <$ 4 and $p_{T} \geq$ 0 GeV/$c$ measured by ALICE in PbPb collisions at $\sqrt{s_{NN}}$ = 2.76 TeV \cite{alice2}. The centrality integrated $R_{AA}$ is 0.545 $\pm$ 0.032(stat.) $\pm$ 0.084(syst.), which clearly indicates the $J/\psi$ suppression relative to the $pp$ data. The contribution from B-meson decays to the inclusive $J/\psi$ yield in the ALICE acceptance has been estimated using the LHCb measurements in $pp$ collisions at $\sqrt{s}$ = 7 TeV. The LHCb collaboration estimated the $b$ fraction of the inclusive $J/\psi$ yield to be about 10\% in $pp$ collisions at 7 TeV. Assuming the same fraction in PbPb at 2.76 TeV, ALICE has found that the difference in the $R_{AA}$ results between the prompt and inclusive $J/\psi$ does not exceed 11\%. 

For comparison, Fig.~\ref{fig:alice_raa} also displays $R_{AA}$ of $J/\psi$ from PHENIX in AuAu collisions at $\sqrt{s_{NN}}$ = 200 GeV \cite{phenix2}. The nuclear modification factor $R_{AA}$ for inclusive $J/\psi$ from ALICE is almost a factor of three larger than the PHENIX data at forward rapidity (1.2 $< |y| <$ 2.2) for $N_{\mathrm{part}} \geq$ 180. In addition, differently from the PHENIX results, the ALICE data do not show centrality dependence.

\section{Comparison with Models}\label{sec:model}

\begin{figure}[t!]
\centering
\includegraphics[width=15cm]{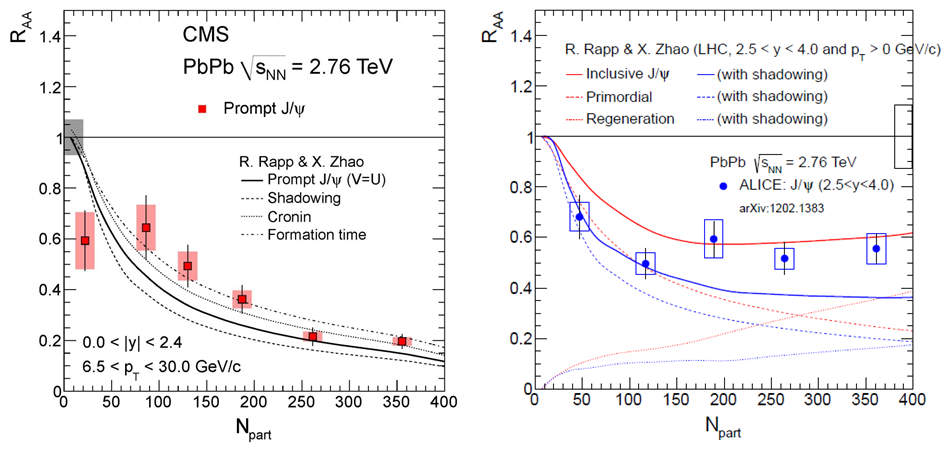}
\caption{Comparison of the experimental data on $R_{AA}$ of $J/\psi$ with the transport model calculations by Zhao and Rapp as a function of $N_{\mathrm{part}}$ in PbPb collisions at $\sqrt{s_{NN}}$ = 2.76 TeV \cite{zhao1}. The shadowing, Cronin, and formation-time effects are taken into account for the prompt $J/\psi$ production at midrapidity on the left panel. The recombined $J/\psi$s are important at low $p_{T}$ for ALICE on the right panel, but negligible at high $p_{T}$ for CMS.}
\label{fig:raa_zhao}
\end{figure}
 
There have been great efforts to understand the $J/\psi$ production in PbPb collisions at LHC. Figure~\ref{fig:raa_zhao} shows the comparison of the experimental data on $R_{AA}$ of $J/\psi$ with one of the transport model calculations as a function of centrality. This very transport model is based on the Boltzmann equation with the spectral properties of charmonia constrained by the thermal lattice QCD \cite{zhao1}. The shadowing effect in the incoming nuclei has been introduced by artificially lowering the $c\bar{c}$ cross section, $d\sigma_{c\bar{c}}/dy$, in this (as well as other) transport model(s). As indicated in the left panel of Fig.~\ref{fig:raa_zhao}, the shadowing effect somewhat reduces the prompt $J/\psi$ yields in the CMS acceptance (6.5 $< p_{T} <$ 30.0 GeV/$c$ in $|y| <$ 2.4), but the Cronin and formation-time effects are shown to enhance them. In the end, the model can reproduce the experimental data reasonably well except the most peripheral bin. On the other hand, recombination becomes the most important process, especially, at the low-$p_{T}$ region. The right panel of Fig.~\ref{fig:raa_zhao} indicates that the recombination component can account for more than half of the inclusive $J/\psi$ yields in the ALICE acceptance ($p_{T} >$ 0 GeV/$c$ in 2.5 $< y <$ 4). It is noteworthy to point out that the ALICE data favor the existence of the shadowing effect in peripheral collisions, but disfavor it in central collisions. 

\begin{figure}[t!]
\begin{center}
\includegraphics[width=9cm]{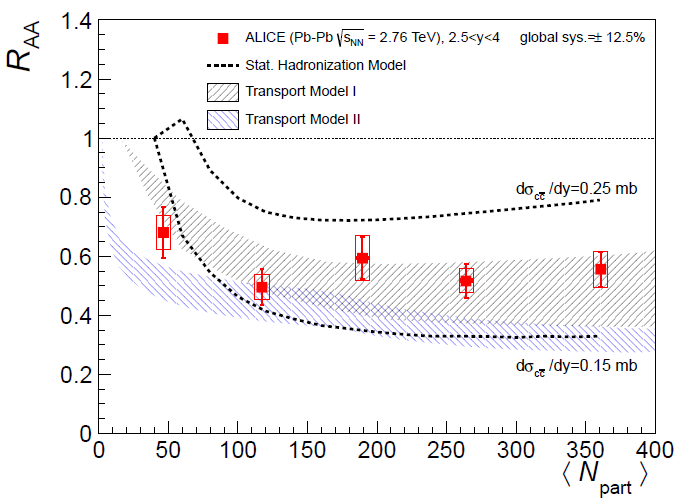}
\caption[]{
Comparison of the ALICE data on $R_{AA}$ of inclusive $J/\psi$ with various model calculations as a function of $N_{\mathrm{part}}$ in PbPb collisions at $\sqrt{s_{NN}}$ = 2.76 TeV \cite{alice2}. The hatched areas represent the two transport model calculations \cite{zhao1,liu1}. (The upper hatched area represents the same range as indicated by the two solid lines by Zhao and Rapp in the right panel of Fig.~\ref{fig:raa_zhao}.) The dashed lines represent the calculations by the statistical hadronization model \cite{anton1}.} 
\label{fig:raa_models}
\end{center}
\end{figure}
 
Figure~\ref{fig:raa_models}, taken from Ref.~\cite{alice2}, shows the comparison of the ALICE data with another transport model calculation that again employs the Boltzmann-type transport equation with the suppression and regeneration parameters determined by the hydrodynamic equation \cite{liu1}. (Note that the upper hatched area indicates the same range represented by the two solid lines in the right panel of Fig.~\ref{fig:raa_zhao}.) The lower and upper limits of each hatched area in Fig.~\ref{fig:raa_models} represent with and without, respectively, shadowing. In both transport models, the regenerated $J/\psi$ component amounts about 50\% of the measured inclusive $J/\psi$ yield. 
 
In addition, the dashed lines in Fig.~\ref{fig:raa_models} represent the calculations by the statistical hadronization model that assumes deconfinement and thermal equilibrium of the bulk of the $c\bar{c}$ pairs \cite{anton1}. In this model, charmonium can be formed at the phase boundary by statistical hadronization of $c$ and $\bar{c}$. In Fig.~\ref{fig:raa_models}, the calculations for two $d\sigma_{c\bar{c}}/dy$ values (0.15 and 0.25 mb) are given. Similar with the transport models, it is found that the statistical hadronization model is also sensitive to the $c\bar{c}$ cross section in medium. 

\begin{figure}[t!]
\centering
\includegraphics[width=15cm]{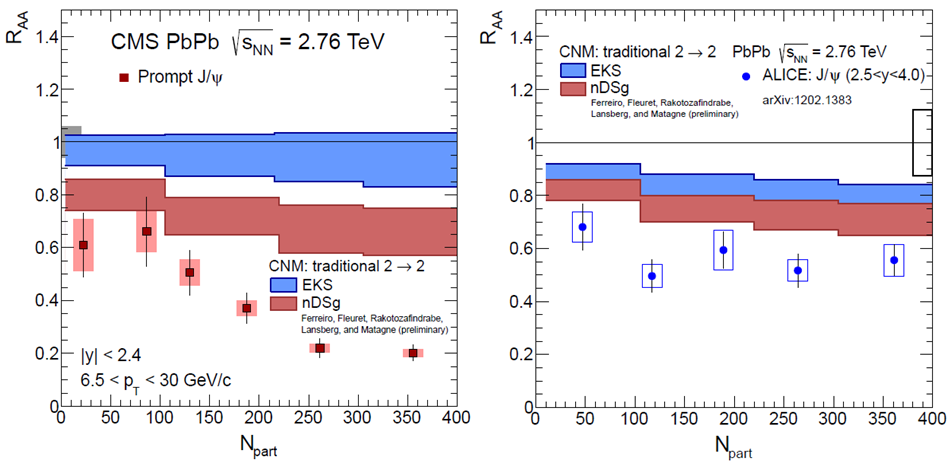}
\caption{Comparison of the CMS (left) and ALICE (right) data on $R_{AA}$ of $J/\psi$ with the QCD calculations for the traditional 2 $\rightarrow$ 2 kinematics at a parton level in PbPb collisions at $\sqrt{s_{NN}}$ = 2.76 TeV. The blue (red) band was obtained by using the EKS (nDSg) parameterization for the modified nuclear parton distribution functions \cite{ferreiro1}.}
\label{fig:comp_22}
\end{figure}

Presently, a more detailed study for the cold nuclear matter effects on the $J/\psi$ production in heavy-ion collisions is underway using different parametrizations of the nuclear parton distribution functions (nPDF). (Some examples of nPDF can be found in \cite{eskola1}.) Figure~\ref{fig:comp_22} shows the comparison of the CMS and ALICE data on $R_{AA}$ of $J/\psi$ with the QCD calculations in the traditional 2 $\rightarrow$ 2 processes at a parton level ($g + g \rightarrow J/\psi + g$) in PbPb collisions at $\sqrt{s_{NN}}$ = 2.76 TeV \cite{ferreiro1}.

\begin{figure}[t!]
\centering
\includegraphics[width=15cm]{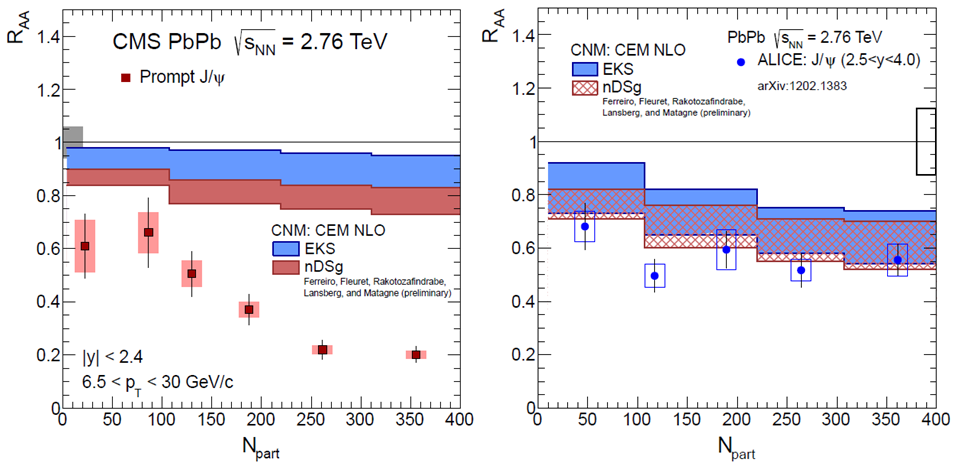}
\caption{Same as Fig.~\ref{fig:comp_22}, but with the calculations obtained via the next-to-leading order of the color evaporation model \cite{ferreiro1}.}
\label{fig:comp_cem}
\end{figure}

In addition, Fig.~\ref{fig:comp_cem} presents the comparison of the LHC data with the calculations obtained by the next-to-leading order (NLO) of the color evaporation model (CEM) before $k_{T}$ smearing \cite{ferreiro1}. The overall cold nuclear matter effect caused by the modification of nPDF underestimates the suppression of the $J/\psi$ yields relative to the $pp$ data. (Only the nDSg parametrization in NLO of CEM barely touches the ALICE data at forward region.) The pPb run at LHC, being planned at the beginning of 2013, will be crucial to scrutinize the cold nuclear matter effects from hot final state effects in the heavy-ion data at LHC. 

\section{$\Upsilon$ from CMS}\label{sec:upsilon} 

Thanks to higher beam energy and luminosity, $pp$ as well as heavy-ion collisions at LHC provide a plenty of $b$ quarks and, thus, bottomonia for the study of QGP. Although bottomonia are not the subjects of this workshop, we summarize the results on the $\Upsilon$ family, as the physics is quite relevant to the charmonium production in heavy-ion collisions. 
 
\begin{figure}[t!]
\begin{center}
\includegraphics[width=15cm]{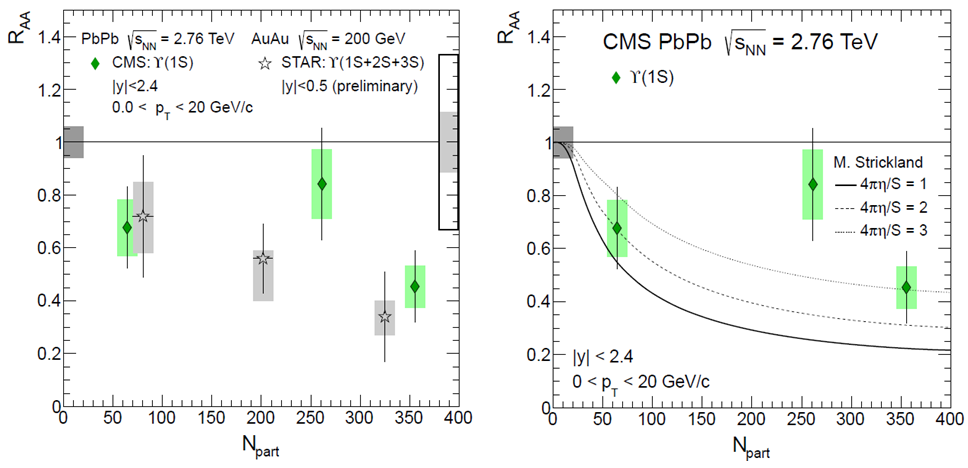}
\caption[]{
Nuclear modification factor $R_{AA}$ of $\Upsilon$(1$S$) as a function of $N_{\mathrm{part}}$ in PbPb collisions at $\sqrt{s_{NN}}$ = 2.76 TeV from CMS \cite{cms1}. The measured $\Upsilon$s are for 0 $< p_{T} <$ 20 GeV/$c$ in $|y| <$ 2.4. The grey box at $R_{AA}$ = 1 and $N_{\mathrm{part}}$ = 0 represents a global scale uncertainty of 6\% originating from the integrated luminosity of the $pp$ data sample. The statistical and systematic uncertainties are shown by bars and boxes, respectively. (Left) The CMS data are in comparison with the $\Upsilon$(1$S$+2$S$+3$S$) data measured by STAR ($|y| <$ 0.5) in AuAu collisions at $\sqrt{s_{NN}}$ = 200 GeV \cite{star3}. The light-grey and open boxes at $N_{\mathrm{part}}$ = 400 represent the systematic and statistical, respectively, uncertainties in $pp$ collisions for STAR. (Right) The CMS data are compared with the theoretical calculations by the anisotropic hydrodynamic model with different options for the ratio of plasma shear viscosity to entropy density $4\pi\eta/S$ \cite{strick1}. See the text for the details.} 
\label{fig:uraa_npart}
\end{center}
\end{figure}

Figure~\ref{fig:uraa_npart} displays the nuclear modification factor $R_{AA}$ of $\Upsilon$(1$S$) as a function of $N_{\mathrm {part}}$ in PbPb collisions at $\sqrt{s_{NN}}$ = 2.76 TeV from CMS \cite{cms1}. For the most central class (0 - 10\%), $R_{AA}$ of $\Upsilon$(1$S$) is 0.45 $\pm$ 0.14(stat.) $\pm$ 0.08(syst.) $\pm$ 0.03(global) for 0 $< p_{T} <$ 20.0 GeV/$c$, which indicates a suppression by about a factor two. However, large statistical and systematic uncertainties for less central event bins prevent drawing any definite conclusions on the centrality dependence. In the left panel of Fig.~\ref{fig:uraa_npart}, the CMS data are compared to the preliminary STAR data for $\Upsilon$(1$S$+2$S$+3$S$) in AuAu collisions at $\sqrt{s_{NN}}$ = 200 GeV \cite{star3}. It is observed that a suppression of inclusive $\Upsilon$s at RHIC is similar to that at LHC despite the large gap in beam energies. 

In the right panel of Fig.~\ref{fig:uraa_npart}, the CMS data are compared to the calculations by the anisotropic hydrodynamic model (AHYDRO) with different options for the ratio of the plasma shear viscosity to the entropy density ($4\pi\eta/S$) \cite{strick1}, where the term `anisotropy' implies local momentum-space anisotropy due to finite plasma shear viscosity. The theoretical calculations for $4\pi\eta/S$ = 1 $\sim$ 3 can reproduce the measured centrality dependence of $R_{AA}$ for $\Upsilon$(1$S$). Note that, in the right panel of Fig.~\ref{fig:uraa_npart}, the author of \cite{strick1} has assumed that about 51\% of $\Upsilon$(1$S$) are produced directly and that all excited $\Upsilon$ states have the same $R_{AA}$ as the $\chi_{b1}$. 

\begin{figure}[t!]
\begin{center}
\includegraphics[width=15cm]{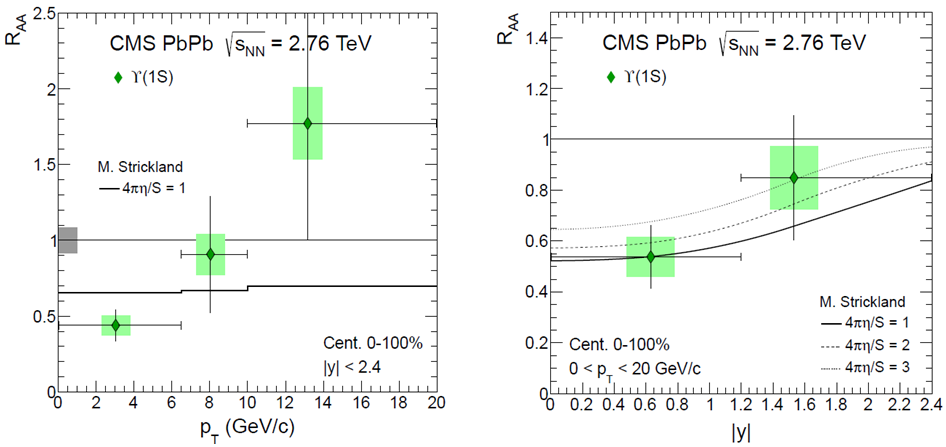}
\caption[]{
Nuclear modification factor $R_{AA}$ of $\Upsilon$(1$S$), integrated over centrality, as functions of $p_{T}$ (left) and $y$ (right) in PbPb collisions at $\sqrt{s_{NN}}$ = 2.76 TeV from CMS \cite{cms1}. $R_{AA}$ is integrated over $|y| <$ 2.4 for the $p_{T}$ dependence and over 0 $< p_{T} <$ 20 GeV/$c$ for the $y$ dependence. The grey boxes at $R_{AA}$ = 1 represent global scale uncertainties of 6\% originating from the integrated luminosity of the $pp$ data sample. The statistical and systematic uncertainties are shown by bars and boxes, respectively. The CMS data are compared to the calculations by the anisotropic hydrodynamic model with different options for the ratio of plasma shear viscosity to entropy density $4\pi\eta/S$ \cite{strick1}. See the text for the details.} 
\label{fig:uraa_pty}
\end{center}
\end{figure}

Figure~\ref{fig:uraa_pty} presents $R_{AA}$ of $\Upsilon$(1$S$) as functions of $p_{T}$ and $y$ in PbPb collisions at $\sqrt{s_{NN}}$ = 2.76 TeV from CMS \cite{cms-u}. For the lowest $p_{T}$ bin from 0 to 6.5 GeV/$c$, $R_{AA}$ of $\Upsilon$(1$S$) is 0.44 $\pm$ 0.10(stat.) $\pm$ 0.06(syst.) $\pm$ 0.04(global) in $|y| <$ 2.4 for the minimum bias condition on centrality, which again indicates a suppression by about a factor of two, but the suppression disappears for $p_{T} >$ 6.5 GeV/$c$. For the rapidity dependence, a slightly larger suppression is observed at midrapidity for minimum bias events, but the large statistical uncertainties do not allow us to make conclusions on the rapidity dependence at the moment. 

In Fig.~\ref{fig:uraa_pty}, the CMS data are again compared with the calculations by AHYDRO with three different values of $4\pi\eta/S$ \cite{strick1}. The calculations for $4\pi\eta/S$ = 1 $\sim$ 3 can reproduce the measured rapidity dependence of $R_{AA}$ for $\Upsilon$(1$S$) reasonably well, but seem to be much flatter than the experimental data in the $p_{T}$ dependence. However, much more precise data are required to draw firm conclusions. 

\begin{figure}[t!]
\centering
\includegraphics[width=9.5cm]{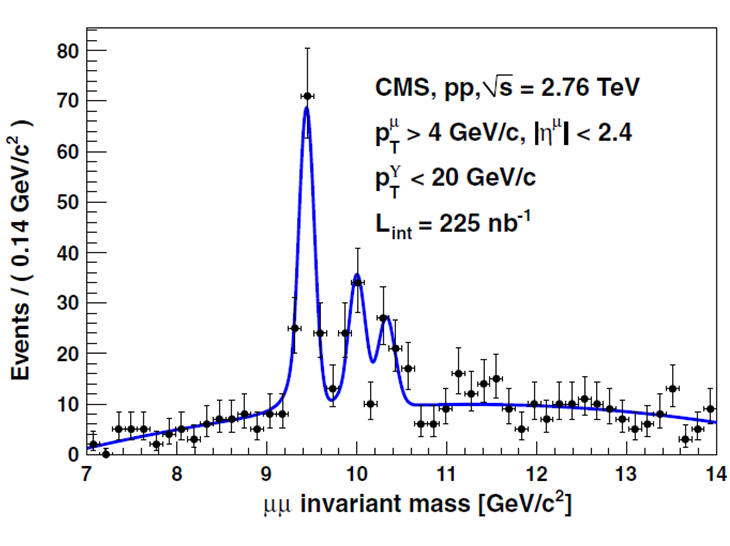}
\includegraphics[width=9.5cm]{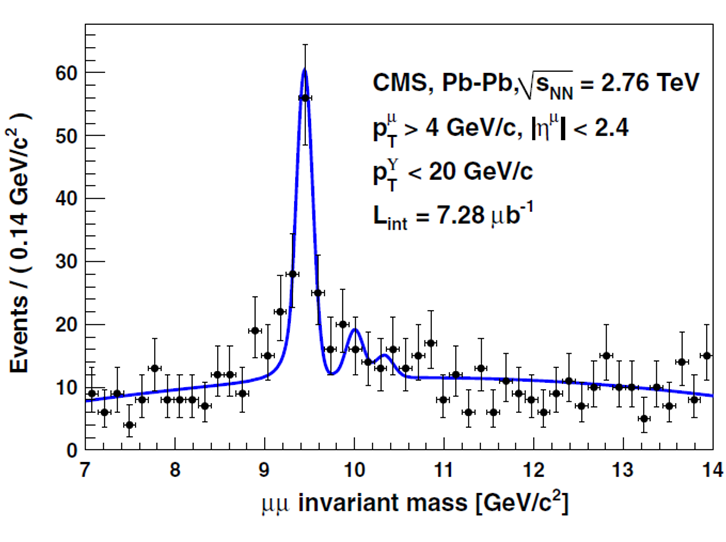}
\caption{
Invariant mass distributions of $\mu^{+}\mu^{-}$ pairs for $p_{T}^{\mu} >$ 4 GeV/$c$ in $pp$ (top) and PbPb (bottom) collisions at $\sqrt{s_{NN}}$ = 2.76 TeV from CMS \cite{cms-u}. The solid lines are the fit functions described in the text.}
\label{fig:u_mass}
\end{figure}
 
Finally, the excellent resolution of the CMS detector for muons allows us to separate $\Upsilon$(1$S$) from the higher $\Upsilon$(2$S$+3$S$) states in heavy-ion collisions. Figure~\ref{fig:u_mass} shows the invariant mass distributions of $\mu^{+}\mu^{-}$ pairs for the $p_{T}$ of each muon, $p_{T}^{\mu}$, larger than 4 GeV/$c$ in $pp$ as well as PbPb collisions at $\sqrt{s_{NN}}$ = 2.76 TeV \cite{cms-u}. Similar with the $J/\psi$ analysis, each $\Upsilon$ state is parameterized by a `crystal ball' function. Since the three $\Upsilon$ states partially overlap in the measured mass range, a simultaneous fit with three `crystal ball' functions added to the second-order polynomial function for the background has been performed. The solid lines in Fig.~\ref{fig:u_mass} represent the final results of fitting. 
 
The comparison of the $\mu^{+}\mu^{-}$ invariant mass distribution in $pp$ with that in PbPb collisions clearly demonstartes that the higher $\Upsilon$(2$S$+3$S$) states are suppressed relative to the $\Upsilon$(1$S$) state in PbPb collisions. The fit functions to the $\mu^{+}\mu^{-}$ invariant mass distributions for $pp$ and PbPb give a double ratio 
\begin{equation}
{{\Upsilon(2S+3S)/\Upsilon(1S)|_{\mathrm {PbPb}}}\over{\Upsilon(2S+3S)/\Upsilon(1S)|_{pp}}} = 0.31 ^{+0.19}_{-0.15}({\mathrm {stat.}}) \pm 0.03({\mathrm {syst.}}),
\label{eq:double}
\end{equation} 
where the systematic uncertainty is from varying the line shape in the simultaneous fit procedure, thus taking into account partial cancellation of systematic effects. 

In order to check the significance of Eq.~\ref{eq:double}, an ensemble of one million pseudo experiments has been generated with the signal line shape obtained from the $pp$ data (the top panel of Fig.~\ref{fig:u_mass}), the background line shapes from both date sets, and a double ratio of Eq.~\ref{eq:double} equal to unity within the statistical and systematic uncertainties. In this test, the probability to find the measured double ratio of 0.31 or less is estimated to be 0.9\%, which corresponds to the 2.4 $\sigma$ effect in a one-tail integral of a Gaussian distribution. 

Currently, a great number of theoretical models are attempting to understand the $\Upsilon$ production data from CMS. Some of them are successful reproducing the centrality dependence of $R_{AA}$ for $\Upsilon$(1$S$) and the double ratio given in Eq.~\ref{eq:double} \cite{rapp_y,song_y,brez_y}.

\section{Summary}\label{sec:summ}

This article summarizes the experimental data on the $J/\psi$ production via $\mu^{+}\mu^{-}$ channel in PbPb collisions at $\sqrt{s_{NN}}$ = 2.76 TeV, measured by the ALICE, ATLAS, and CMS collaborations at LHC. Soon after the first heavy-ion run at LHC in 2010, ATLAS published the central-to-peripheral ratio of the $J/\psi$ yield, $R_{cp}$, and demonstrated that more suppression occurs for more central collisions. Later CMS published much more detailed results on the $J/\psi$ production using the reference $pp$ data. In the CMS analysis, the prompt and non-prompt $J/\psi$s are separated by fitting the $\mu^{+}\mu^{-}$ invariant mass and the pseudo-proper decay length distributions simultaneously. The prompt and non-prompt $J/\psi$ yields were analyzed in PbPb collisions, and compared to those in $pp$ collisions at the same beam energy in order to learn whether there exists any medium effect.

The prompt $J/\psi$ yield is suppressed by a factor of about 5 for the most central class (0 - 10\%), which diminishes to about 1.6 for less central events (50 - 100\%). The $R_{AA}$ of the prompt $J/\psi$ at LHC is more than a factor two smaller than the $J/\psi$ data at RHIC. 

The $R_{AA}$ distribution of prompt $J/\psi$ displays no $p_{T}$ dependence with the current widths of the $p_{T}$ bins. In addition, slightly more suppression is observed at midrapidity than at forward rapidity, which is opposite to the behavior observed at low $p_{T}$ at RHIC. On the other hand, the non-prompt $J/\psi$s decayed from the B mesons are also suppressed by a factor of about 2.7 with respect to the $pp$ data. This result indicates the energy loss of $b$-quarks in medium for the first time. 

ALICE measured $J/\psi$s at low $p_{T}$ in the forward region that is complimentary to the ATLAS and CMS acceptances. The low-$p_{T}$ $R_{AA}$, integrated over centrality, shows a suppression relative to $pp$ in the forward region. The comparison with the PHENIX data at forward rapidity indicates that $R_{AA}$ of $J/\psi$ by ALICE is almost a factor of three larger than that by PHENIX for $N_{\mathrm{part}} \geq$ 180. Furthermore, the ALICE data do not show the centrality dependence contrary to the PHENIX data. 
 
The $R_{AA}$ distributions of prompt $J/\psi$ are compared with the calculations by the microscopic transport models and the statistical hadronization model. From these comparisons we learned that (1) regeneration is more important at low $p_{T}$, (2) the theoretical calculations are sensitive to the $c\bar{c}$ cross section in medium, and (3) the calculations also depend on the nuclear parton distribution function in the incoming nuclei. Lastly, it has been demonstrated that the modification of nPDF alone cannot satisfactorily describe the LHC data on the $J/\psi$ yield. Presently, more detailed investigation on the cold nuclear matter effects in heavy-ion collisions is underway by testing different parametrizations of nPDF. The upcoming pPb run at LHC will be crucial to scrutinize the cold nuclear matter effects in the heavy-ion data. 

The $\Upsilon$(1$S$) state in PbPb collisions is suppressed by a factor of about 2 in the most central class (0 - 10 \%) when integrated over all $p_{T}$ and $y$. Suppression factor is large at low $p_{T}$, but negligible at high $p_{T}$. The ratio of $\Upsilon$(2$S$+3$S$) to $\Upsilon$(1$S$) in PbPb collisions is suppressed by a factor of about 3 with respect to that in $pp$ for $p_{T}^{\mu} >$ 4 GeV/$c$. The probability to obtain the measured, or smaller, value of the double ratio, if the true suppression factor is unity, has been estimated to be less than 1\%. Several theoretical efforts to understand the $\Upsilon$ production data by CMS are ongoing.

\section*{Acknowledgements}
This work was partially supported by the National Research Foundation of Korea under grant No. K20802011718-11B1301-00610. 
The author is grateful to Anton Andronic for various comments and, in particular, providing the dimuon invariant mass distributions for the ALICE collaboration.


\begin{thebibliography}{99}

\bibitem{qgp1} A.~Bazavov, T.~Bhattacharya, M.~Cheng, N.~H.~Christ, C.~DeTar, S.~Ejiri, S.~Gottlieb and R.~Gupta {\it et al.},
  Phys.\ Rev.\ D {\bf 80}, 014504 (2009)
  [arXiv:0903.4379 [hep-lat]].
\bibitem{matsui1} T.~Matsui and H.~Satz,
  Phys.\ Lett.\ B {\bf 178}, 416 (1986).
\bibitem{satz1} H.~Satz,
  Nucl.\ Phys.\ A {\bf 783}, 249 (2007)
  [hep-ph/0609197].
\bibitem{satz2} H.~Satz,
  J.\ Phys.\ G {\bf 32}, R25 (2006)
  [hep-ph/0512217].
\bibitem{mocsy1} A.~Mocsy,
  Eur.\ Phys.\ J.\ C {\bf 61}, 705 (2009)
  [arXiv:0811.0337 [hep-ph]].
\bibitem{phenix1} A.~Adare {\it et al.}  [PHENIX Collaboration],
  Phys.\ Rev.\ C {\bf 77}, 024912 (2008)
  [Erratum-ibid.\ C {\bf 79}, 059901 (2009)]
  [arXiv:0711.3917 [nucl-ex], arXiv:0903.4845 [nucl-ex]].
\bibitem{sps1} B.~Alessandro {\it et al.}  [NA50 Collaboration],
  Eur.\ Phys.\ J.\ C {\bf 39}, 335 (2005)
  [hep-ex/0412036].
\bibitem{sps2} R.~Arnaldi {\it et al.}  [NA60 Collaboration],
  Phys.\ Rev.\ Lett.\  {\bf 99}, 132302 (2007).
\bibitem{sps3} M.~C.~Abreu {\it et al.}  [NA38 Collaboration],
  Phys.\ Lett.\ B {\bf 449}, 128 (1999).
\bibitem{phenix2} A.~Adare {\it et al.}  [PHENIX Collaboration],
  Phys.\ Rev.\ C {\bf 84}, 054912 (2011)
  [arXiv:1103.6269 [nucl-ex]].
\bibitem{star1} B.~I.~Abelev {\it et al.}  [STAR Collaboration],
  Phys.\ Rev.\ C {\bf 80}, 041902 (2009)
  [arXiv:0904.0439 [nucl-ex]].
\bibitem{phenix3} A.~Adare {\it et al.}  [PHENIX Collaboration],
  Phys.\ Rev.\ Lett.\  {\bf 101}, 122301 (2008)
  [arXiv:0801.0220 [nucl-ex]].
\bibitem{alice2} B.~Abelev {\it et al.}  [ALICE Collaboration],
  arXiv:1202.1383 [hep-ex].
\bibitem{atlas1} G.~Aad {\it et al.}  [ATLAS Collaboration],
  Phys.\ Lett.\ B {\bf 697}, 294 (2011)
  [arXiv:1012.5419 [hep-ex]].
\bibitem{cms1} S.~Chatrchyan {\it et al.}  [CMS Collaboration],
  JHEP {\bf 1205}, 063 (2012)
  [arXiv:1201.5069 [nucl-ex]].
\bibitem{cms-u} S.~Chatrchyan {\it et al.}  [CMS Collaboration],
  Phys.\ Rev.\ Lett.\  {\bf 107}, 052302 (2011)
  [arXiv:1105.4894 [nucl-ex]].
\bibitem{alice-d} K.~Aamodt {\it et al.}  [ALICE Collaboration],
  JINST {\bf 3}, S08002 (2008).
\bibitem{atlas-d} G.~Aad {\it et al.}  [ATLAS Collaboration],
  JINST {\bf 3}, S08003 (2008).
\bibitem{cms-d} S.~Chatrchyan {\it et al.}  [CMS Collaboration],
  JINST {\bf 3}, S08004 (2008).
\bibitem{pythia} 
  T.~Sjostrand, S.~Mrenna and P.~Z.~Skands,
  JHEP {\bf 0605}, 026 (2006)
  [hep-ph/0603175].
\bibitem{hydjet}
  I.~P.~Lokhtin and A.~M.~Snigirev,
  Eur.\ Phys.\ J.\ C {\bf 45}, 211 (2006)
  [hep-ph/0506189].
\bibitem{cdf1} D.~Acosta {\it et al.}  [CDF Collaboration],
  Phys.\ Rev.\ D {\bf 71}, 032001 (2005)
  [hep-ex/0412071].
\bibitem{cms2} V.~Khachatryan {\it et al.}  [CMS Collaboration],
  Eur.\ Phys.\ J.\ C {\bf 71}, 1575 (2011)
  [arXiv:1011.4193 [hep-ex]].
\bibitem{star2} Z.~Tang [STAR Collaboration],
  J.\ Phys.\ G {\bf 38}, 124107 (2011)
  [arXiv:1107.0532 [hep-ex]].
\bibitem{phenix-pty} A.~Adare {\it et al.}  [PHENIX Collaboration],
  Phys.\ Rev.\ Lett.\  {\bf 98}, 232301 (2007)
  [nucl-ex/0611020].
\bibitem{pillot1} P.~Pillot [ALICE Collaboration],
  J.\ Phys.\ G {\bf 38}, 124111 (2011)
  [arXiv:1108.3795 [hep-ex]].
\bibitem{alice1} A.~Andronic, private communication.
\bibitem{zhao1} X.~Zhao and R.~Rapp,
  Nucl.\ Phys.\ A {\bf 859}, 114 (2011)
  [arXiv:1102.2194 [hep-ph]] and private communication. 
\bibitem{liu1} Y.~p.~Liu, Z.~Qu, N.~Xu and P.-f.~Zhuang,
  Phys.\ Lett.\ B {\bf 678}, 72 (2009)
  [arXiv:0901.2757 [nucl-th]].
\bibitem{anton1} A.~Andronic, P.~Braun-Munzinger, K.~Redlich and J.~Stachel,
  J.\ Phys.\ G {\bf 38}, 124081 (2011)
  [arXiv:1106.6321 [nucl-th]].  
\bibitem{eskola1} K.~J.~Eskola, H.~Paukkunen and C.~A.~Salgado,
  JHEP {\bf 0904}, 065 (2009)
  [arXiv:0902.4154 [hep-ph]].
\bibitem{ferreiro1} A.~Rakotozafindrabe, E.~G.~Ferreiro, F.~Fleuret, J.~P.~Lansberg and N.~Matagne,
  Nucl.\ Phys.\ A {\bf 855}, 327 (2011)
  [arXiv:1101.0488 [hep-ph]] and private communication.
\bibitem{star3} H.~Masui [STAR Collaboration],
  J.\ Phys.\ G {\bf 38}, 124002 (2011)
  [arXiv:1106.6021 [nucl-ex]].
\bibitem{strick1} M.~Strickland,
  Phys.\ Rev.\ Lett.\  {\bf 107}, 132301 (2011)
  [arXiv:1106.2571 [hep-ph]].
\bibitem{rapp_y} A.~Emerick, X.~Zhao and R.~Rapp,
  Eur.\ Phys.\ J.\ A {\bf 48}, 72 (2012)
  [arXiv:1111.6537 [hep-ph]].
\bibitem{song_y} T.~Song, K.~C.~Han and C.~M.~Ko,
  arXiv:1112.0613 [nucl-th].
\bibitem{brez_y} F.~Brezinski and G.~Wolschin,
  Phys.\ Lett.\ B {\bf 707}, 534 (2012)
  [arXiv:1109.0211 [hep-ph]].

\end{thebibliography}
\end{document}